**Effect of age on the variability and stability of gait: a cross-sectional treadmill study in healthy individuals between 20 and 69 years of age**

**Authors:**

Philippe Terrier[ab§], PhD

Fabienne Reynard[b], PT MSc

[a] IRR, Institute for Research in Rehabilitation, Sion, Switzerland

[b] Clinique romande de réadaptation SUVACare, Sion, Switzerland

**§ Corresponding author:**
Dr Philippe Terrier
Clinique romande de réadaptation SUVACare
Av. Gd-Champsec 90
1951 Sion
Switzerland
Tel.: +41-27-603-23-91
E-mail: **Philippe.Terrier@crr-suva.ch**

**Abstract**

Falls during walking are a major health issue in the elderly population. Older individuals are usually more cautious, walk more slowly, take shorter steps, and exhibit increased step-to-step variability. They often have impaired dynamic balance, which explains their increased falling risk. Those locomotor characteristics might be the result of the neurological/musculoskeletal degenerative processes typical of advanced age or of a decline that began earlier in life. In order to help determine between the two possibilities, we analyzed the relationship between age and gait features among 100 individuals aged 20–69. Trunk acceleration was measured during a 5-min treadmill session using a 3D accelerometer. The following dependent variables were assessed: preferred walking speed, walk ratio (step length normalized by step frequency), gait instability (local dynamic stability, Lyapunov exponent method), and acceleration variability (root mean square [RMS]). Using age as a predictor, linear regressions were performed for each dependent variable. The results indicated that walking speed, walk ratio and trunk acceleration variability were not dependent on age ($R^2<2\%$). However, there was a significant quadratic association between age and gait instability in the mediolateral direction ($R^2=15\%$). We concluded that most of the typical gait features of older age do not result from a slow evolution over the life course. On the other hand, gait instability likely begins to increase at an accelerated rate as early as age 40–50. This finding supports the premise that local dynamic stability is likely a relevant early indicator of falling risk.

**Keywords:** Human locomotion; Accelerometer; Dynamic balance; Fall risk; Nonlinear analysis; Local dynamic stability; Acceleration RMS

# 1. Introduction

Falls during walking are a major health issue in older adults. Elderly individuals exhibit more conservative gait patterns characterized by slower preferred walking speeds (PWS) and reduced step lengths [1], which are indications of greater cautiousness [2]. Musculoskeletal weakness is strongly associated with falls [3]. The decline of cognitive function is correlated with fall risk [4] and is specifically associated with reduced walking speed [5].

Many different methods have been proposed to describe gait characteristics in the older population to determine the causes of falls. Besides basic spatiotemporal gait features that are modified in older, healthy adults compared to their younger counterparts [1], it is also important to assess the variability of the gait pattern, which is caused by the decreased ability to optimally control gait from one stride to the next [6]. In this context, the root mean square (RMS) of trunk acceleration is often used as a measure of gait variability [7]. Optimal dynamic balance results in smooth trunk acceleration during walking; therefore, a low RMS value is considered evidence of a healthy gait. Another popular method is the estimation of local dynamic stability (LDS), which is derived from chaos theory (maximal Lyapunov exponent [8]). This method takes the nonlinear features of human movement into account more effectively than classical variability estimates (RMS, standard deviation, coefficient of variation). It is assumed that motor control ensures a dynamically stable gait (high LDS) if the divergence remains low between trajectories in a reconstructed state space that reflects gait dynamics. The usefulness of gait LDS to assess gait stability and falling risk has been shown in theoretical [9], experimental [10], and clinical [11] studies [12].

Although the abovementioned parameters have already been proposed to characterize gait in elderly individuals [2, 13, 14], there is insufficient information regarding the changes in these parameters with age. Most studies have compared older adults to matched young controls. However, some aspects of cognitive capabilities decline as early as the second or third decades of life [15]. Similarly, significant strength loss in the lower extremities begins between ages 40 and 50 [16]. Because musculoskeletal and cognitive status are key factors in the etiology of falls in the elderly, gait features in middle-aged adults (40–60 y) demand further investigation [7]. In other words, it is unclear whether the idiosyncrasy of gait in elderly individuals is the result of musculoskeletal/neurological degenerative processes that occur with advanced age, or whether it is the result of a slower evolution throughout the life course.

The objective of the present cross-sectional study, therefore, was to document the effect of age on gait features in 100 healthy individuals aged 20–69. In addition to basic spatiotemporal measures (PWS, step length), gait variability (RMS) and gait stability (LDS) were analyzed while each participant walked on a treadmill. More generally, we sought to assess the extent to which the typical gait characteristics observed in older adults were already present in middle-aged individuals.

## 2. Methods

*2.1 Subjects*

The study included 100 healthy subjects (50 males, 50 females) without neurological or orthopedic conditions. There were 10 males and 10 females for each decade between the ages of 20 and 69. Their anthropometric features are presented in Table 1. All participants were accustomed to treadmill walking. A subset (95/100) of the subjects was analyzed in a parallel study about LDS reliability [17]. The study was approved by the regional medical ethics committee (Commission Cantonale Valaisanne d'Ethique Médicale, Sion, Switzerland).

*2.2. Experimental procedure and data pre-processing*

The subjects wore a tri-axial accelerometer (Physilog® System, Gaitup, Lausanne, Switzerland) fixed with a belt at the anterior upper trunk level, 5 cm under the sternal notch. The accelerometer measured trunk acceleration along 3 axes: mediolateral (ML), vertical, (V), and anteroposterior (AP). Each participant walked barefoot on a treadmill (Venus model, h/p/cosmos®, Traunstein, Germany) while wearing a safety harness that did not impede movement of the arms and legs. PWS was assessed using the method described by Dingwell and Marin [18]. Trunk accelerations were recorded for 5 min while the subjects walked at PWS. Because acceleration data had already been used in the above-mentioned study [17], we employed an identical method for consistency. The data analysis was performed with MATLAB R2013a (MathWorks, Natick, MA). To lower the effect of sensor misplacement, the 3D-acceleration signals were reoriented according to the procedure proposed by Moe-Nilssen [17, 19]. To avoid starting effects, the first 5 s were discarded. The raw 200-Hz signals were then downsampled to 50 Hz to facilitate the subsequent analyses. Step frequency (SF) was assessed using the fast Fourier transform of the vertical acceleration signal. In the frequency domain, the SF was defined as the higher peak in the 0.5–2.5 Hz band. A duration corresponding to 175 strides was then selected for further analysis (i.e., 152–235 s, depending on individual walking speed and cadence). This length was chosen because it provided sufficient reliability for estimating the LDS and RMS [17, 20].

*2.3 Walk ratio*

The average step length (SL) of the 175 strides was computed from the average treadmill speed (SL=PWS/SF). The walk ratio (WR) is the SL normalized by SF (WR=SL/SF): WR represents what would be SL assuming a SF of 1 step/s. This method is an appropriate means of characterizing gait pattern [21] and takes advantage of the invariant relationship between SL and SF, regardless of walking speed [21].

*2.4 Gait variability (RMS$_{RATIO}$)*

Because acceleration RMS is highly correlated with walking speed [7], the normalization method recently introduced by Sekine [22] was employed. To compute the RMS ratio of the trunk acceleration (RMS$_{RATIO}$) the vector norm (*L*) of the 3D acceleration (*x*, *y*, *z*) for each sample *n* was first computed ($L_n = \sqrt{x_n^2 + y_n^2 + z_n^2}$). The RMS of the vector norm was $L_{RMS} = \sqrt{\frac{1}{N}\sum_{n=1}^{N}(L_n)^2}$. The same procedure was applied to the ML acceleration signal to compute ML$_{RMS}$. The RMS$_{RATIO}$, $RMS_{RATIO} = \frac{ML_{RMS}}{L_{RMS}}$, quantified the proportion of trunk acceleration variability that occurred in the ML direction compared to the total acceleration variability.

*2.5 Gait instability (local dynamic stability)*

The LDS quantification was based on the maximal Lyapunov exponent method using Rosenstein's algorithm. (The reader interested in a full theoretical background may refer to our recently published articles [8, 17] that include a more detailed methodology.) The acceleration signals were time-normalized to a uniform length of 10,000 samples to thwart the trend toward a spurious lower stability in longer signals [17]. The high-dimension attractor was constructed using a uniform time delay of 6 samples for all signals according to the average results of an average mutual information (AMI) analysis. A constant dimension of 6 was selected in accordance with the average results of a global false nearest neighbors (GFNN) analysis. The LDS was estimated from divergence exponents representing the average rates of logarithmic divergence between trajectories located downstream of the nearest neighbors in the attractor. The short-term divergence exponent ($\lambda$) over 1 step was used, and was determined based on a constant number of 29 samples in the divergence diagram [8, 17].

*2.6 Statistics*

The measured variables were age (y), body weight (kg) and height (m), PWS (m·s$^{-1}$), WR (m·Hz$^{-1}$), gait variability ($RMS_{RATIO}$, dimensionless), and gait instability in ML, V, and AP directions (LDS, divergence exponents $\lambda$). Means and standard deviations (SD) by age ranges are presented in Table 1. One-way ANOVAs were performed to estimate the differences among age categories. *F* and *p*-values were reported, as well as $\omega^2$, which was the unbiased version of $\eta^2$ reflecting the proportion of variance explained by group membership. The 95% confidence intervals (CIs) of $\omega^2$ were computed by bootstrapping (5,000 resamples).

A regression analysis was performed for each gait parameter with age as the predictor. In a preliminary analysis, we excluded the existence of a significant relationship between age and body mass or body height, which might have biased the results (not shown). The stepwise regression method indicated that a quadratic model fit better with the LDS data (i.e., LDS = a + $b_1$·age + $b_2$·age$^2$) than the simple linear model ($\hat{y}$ = a + $b_1$·age), which was adopted for the other dependent variables. Based on the inspection of normal probability plots of the residuals, outliers were discarded from the final models. The results are shown in Table 2, with ANOVA analyses that tested whether the regression model was compatible with the assumption of a constant model: the null hypothesis was that all of the regression coefficients were equal to zero. Unbiased coefficients of determination ($\rho^2 \sim R^2$) were calculated using the Olkin-Pratt formula, which reflects the proportion of variance explained by the regression model. The precision of $\rho^2$ was determined with 95% CI using bootstrapping (5,000 resamples). In Figures 1 and 2, scatter plots and best-fit lines illustrate the results of the regression analyses.

## 3. Results

When differences among age categories were considered (Table 1), no significant effects were observed in the anthropometric characteristics (body weight and height). Similarly, spatiotemporal parameters (PWS, WR) were equal among groups; that is, the variance explained by group membership was 0%. Given the 95% CIs, which were below 18%, it is very unlikely that a substantial age effect exists at the population level. The same conclusion can be drawn regarding gait variability (acceleration RMS), as no significant effect was observed ($\omega^2$=3%). While gait stability measured in the AP and V directions were equal among age categories ($\omega^2 \leq 1$%), LDS in the ML direction exhibited a significant difference between groups (p=0.02). Group membership explained 8% of the total variance (95% CI, 1–28%).

    The linear regressions with age as predictors (Table 2 and Figs 1 and 2) confirmed the ANOVA results (Table 1). The only significant age effect was for LDS measured in the ML direction (p<0.001). The quadratic regression model explained 15% of the variance among

individuals (95% CI: 2%–29%). For the other variables, the age effect was negligible, with $\rho^2$ below 4%.

## 4. Discussion

The rationale behind the design of the present study was that (1) there is a strong relationship between musculoskeletal/cognitive deficits and fall risk in older adults, (2) the decline in muscle strength and some aspects of cognitive performance has been reported in middle-aged adults, and (3) there is a lack of studies analyzing the gait of middle-aged individuals. Thus, using trunk accelerometry, we conducted a cross-sectional analysis of 100 individuals of various ages (Table 1) while they walked at PWS on a treadmill. The results show that basic spatiotemporal parameters (PWS, WR) did not change with age (Tables 1 and 2). Acceleration RMS ($RMS_{RATIO}$) also remained constant. Conversely, the mediolateral dynamic stability (ML-LDS) (a valid index of fall risk [12]) findings indicated that increased gait instability might begin as early as the fourth or fifth decade of life (Fig 2).

Several studies have reported lower walking speeds accompanied by increased falling risk in elderly individuals. For instance, in a large longitudinal study (n=7,575), slower gait speed was identified as an important factor in risk of hip fracture [23]. In addition, poorer performance in cognitive tasks was associated with slower gait, suggesting a relationship between cognitive and motor declines [5]. In a recent meta-analysis of 17 studies, Iosa et al. [7] showed that a gradual decline in walking speed likely occurs throughout the life course. However, it is worth noting that no study reported results in individuals in their forties and fifties. Therefore, it is unclear when the decline in walking speed begins. Our results did not reveal a significant decline in PWS with age (Table 2 and Fig 1). We found a change of -0.07 km/h/decade, which was very small compared to the substantial between-subjects variability (SD=0.65 km/h). It is likely that speed decline occurs at a higher rate after age 60, which could explain the differences between young and elderly individuals reported in the literature [7].

Smaller steps are a sign of greater cautiousness. For instance, when healthy young adults walked with their eyes closed, they reduced SL more than SF, which resulted in a lower WR [24]. Similarly, among individuals walking in an artificially destabilized environment (lateral random movement of the walking surface), the cautious gait strategy induced a large decrease in WR [10]. Shorter SLs have been reported in elderly individuals, in accordance with the hypothesis that elderly people walk more carefully [1]. In this study, we did not observe changes in WR (Tables 1 and 2, Fig. 1). The ratio between SL and cadence remained constant, which may indicate that the degree of cautiousness does not evolve with age among young and middle-aged individuals.

The use of acceleration RMS to analyze gait variability has been widely reported [7]. However, the quadratic relationship between acceleration RMS and walking speed has made it difficult to compare individuals with various PWSs [7]. To address this problem, several normalization methods have been proposed [7, 22, 25]. In the present study, we used the normalization method recently advocated by Sekine and colleagues [22]; that is, the $RMS_{RATIO}$ describes the proportion of trunk acceleration variability that occurs in the ML direction compared to total acceleration variability. Other authors have proposed a very similar method [25]. Sekine showed that the $RMS_{RATIO}$ discriminated between hemiplegic patients and healthy controls, with patients exhibiting higher values [22]. Because the lateral balance is important for the maintenance of optimal dynamic stability [26], it is likely that a deficit in balance control induce a higher variability in that direction which is adequately detected by $RMS_{RATIO}$. In the same study, a U-shaped dependency between speed and $RMS_{RATIO}$ was observed, with the minimum corresponding to PWS. This showed that walking may be optimally stable (and hence lower in variability) at PWS.

In the abovementioned meta-analysis by Iosa [7], there was no clear trend in the evolution of RMS with age in adults due to the absence of data for middle-aged individuals and the substantial discrepancies among studies. In our results, $RMS_{RATIO}$ did not change with age. Thus, it could be that healthy adults keep a constant level of gait variability throughout the life course until reaching the sixth decade.

The only significant result in gait stability was found along the ML axis in the current study. However, the other axes (especially the vertical axis) revealed quadratic dependencies with age (Fig. 2) that were not significant due to the substantial between-subjects variability. This result was not surprising. We [17, 27] and others [14] have advocated the use of ML-LDS to assess dynamic balance. Theoretical and experimental results have highlighted the importance of the frontal plane in the regulation of dynamic balance. It is thought that gait is constitutively stable in the AP and V directions, while active control is required to stabilize the body laterally [26]. Furthermore, ML-LDS exhibits better repeatability in both treadmill [17] and overground walking [27]. Finally, there is evidence that ML-LDS can optimally discriminate between healthy and fall-prone individuals [27, 28]. In other words, it is very likely that ML-LDS is more sensitive in detecting deficits in balance control compared to LDS measured in other directions.

Some studies have compared the LDS of young individuals and older adults. Very recently, Bruijn and colleagues [14] compared the dynamic stability of young adults (n=15, age 22) to that of older adults (n=25, age 71). They found significantly more unstable gaits in elderly individuals (ML-LDS of the pelvis, 10% difference). In 2009, Kang and Dingwell [13] highlighted substantial differences between older adults (n=18, age 72) and members of a matched, younger control group (n=17, age 23). Trunk LDS differed by about 40% (graphical estimation). In 2008, Kyvelidou and colleagues [29] showed similar results by comparing 10 young females (aged 24) to 10 older females (aged 73). In their study, the LDS measured at the ankle, knee and hip levels differed by 16, 17, and 10%, respectively. However, there was no statistical significance, which was likely due to the small sample size. In the present study, the quadratic model predicted that ML-$\lambda$ would be 13% higher (i.e., lower stability) at the age of 75 than at the age of 25, in line with the studies cited herein.

## 5. Conclusion

The present study showed that the documented difference between young and elderly individuals [13, 14, 29] is likely the result of an accelerating increase in gait instability that begins earlier in life — possibly as early as the fourth decade (Fig. 2). In contrast, we did not observe significant changes with age in other parameters (speed, SL, and gait variability). Interestingly, a large-scale epidemiological study showed that the frequency of falls in middle-aged adults (46–65 y) was higher than in younger adults (20–45 y) [30]. This confirms that LDS may be a valid index of falling risk [11, 12]. Although the underlying causes of these premature declines in gait stability remain to be investigated, we suspect changes in muscle strength and/or motor-control capabilities. Future longitudinal studies following individuals over many years, before they reach an advanced age, should be conducted to confirm whether LDS is a valid method for the early identification of fall-prone individuals.

**Conflicts of interest statement**

There are no known conflicts of interest.

| N=100 | Global mean | 20–29 y (N=20) | 30-39 y (N=20) | 40-49 y (N=20) | 50-59 y (N=20) | 60-69 y (N=20) | F (4,95) | p | $\omega^2$ |
|---|---|---|---|---|---|---|---|---|---|
| Age (y) | 44.2 (14.1) | 24.7 (2.8) | 34.6 (2.8) | 43.9 (2.9) | 54.8 (2.7) | 63.3 (3.2) | - | - | - |
| Body weight (kg) | 70.2 (14.6) | 68.4 (11.9) | 65.4 (12.8) | 74.2 (15.6) | 71.1 (14.4) | 72.0 (17.2) | 1.10 | 0.36 | 0.00 (0 – 0.15) |
| Body height (m) | 1.72 (0.07) | 1.74 (0.06) | 1.70 (0.08) | 1.74 (0.06) | 1.71 (0.08) | 1.69 (0.06) | 1.94 | 0.11 | 0.04 (0 – 0.19) |
| Preferred walking speed (m·s$^{-1}$) | 1.09 (0.18) | 1.10 (0.15) | 1.13 (0.13) | 1.11 (0.17) | 1.04 (0.24) | 1.06 (0.17) | 0.95 | 0.44 | 0 (0 – 0.18) |
| Walk ratio (m·Hz$^{-1}$) | 0.31 (0.04) | 0.32 (0.04) | 0.32 (0.04) | 0.32 (0.04) | 0.30 (0.05) | 0.30 (0.04) | 0.78 | 0.54 | 0 (0 – 0.12) |
| Gait variability (RMS$_{RATIO}$) | 0.48 (0.07) | 0.49 (0.06) | 0.47 (0.05) | 0.45 (0.07) | 0.47 (0.06) | 0.51 (0.10) | 1.79 | 0.14 | 0.03 (0 – 0.20) |
| ML gait instability (LDS, $\lambda$) | 0.86 (0.06) | 0.85 (0.06) | 0.84 (0.06) | 0.86 (0.04) | 0.88 (0.07) | 0.90 (0.06) | 3.23 | **0.02** | 0.08 (0.01 - 0.28) |
| V gait instability (LDS, $\lambda$) | 1.23 (0.15) | 1.22 (0.18) | 1.19 (0.14) | 1.22 (0.16) | 1.24 (0.13) | 1.26 (0.16) | 0.52 | 0.72 | 0 (0 – 0.12) |
| AP gait instability (LDS, $\lambda$) | 1.09 (0.12) | 1.08 (0.10) | 1.06 (0.11) | 1.12 (0.16) | 1.07 (0.12) | 1.12 (0.12) | 1.12 | 0.35 | 0.01 (0 – 0.16) |

*Table 1. Descriptive statistics and ANOVAs. The 100 participants were classified into 5 age categories and walked for 5 min on a treadmill at preferred walking speed. Trunk accelerations in the mediolateral (ML), vertical (V), and anteroposterior (AP) directions were recorded by a 3D accelerometer. Walk ratio was defined as step length divided by step frequency. Gait variability is the RMS of the lateral acceleration normalized (RMS$_{RATIO}$) to attenuate the influence of speed (see Methods section). Gait instability (local dynamic stability, LDS) was computed using the maximal finite-time Lyapunov exponent method. Mean (SD) is shown for each age category. One-way ANOVAs were used to assess the differences*

*among age categories. F and p-values are shown, as well as $\omega^2$, which is an unbiased equivalent of $\eta^2$; 95% CIs of $\omega^2$ are shown parenthetically.*

| | N | Intercept | Slope | Quadratic term | RMSE | F | p | $\rho^2$ |
|---|---|---|---|---|---|---|---|---|
| **Preferred walking speed (m·s$^{-1}$)** | 99 | 1.17 (0.05) | -0.002 (0.001) | - | 0.16 | 3 | 0.09 | 0.02 (0 – 0.11) |
| **Walk ratio (m·Hz$^{-1}$)** | 99 | 0.33 (0.01) | -0.0004 (0.0003) | - | 0.04 | 1.7 | 0.19 | 0.01 (0 – 0.10) |
| **Gait variability (RMS$_{RATIO}$)** | 99 | 0.47 (0.022) | 0.0000 (0.0004) | - | 0.07 | 0.01 | 0.93 | 0 (0 – 0) |
| **ML gait instability (LDS, λ)** | 99 | 0.92 (0.06) | -0.005 (0.0027) | 0.00007 (0.00003) | 0.05 | 9.4 | **<0.001** | 0.15 (0.02 – 0.29) |
| **V gait instability (LDS, λ)** | 100 | 1.46 (0.16) | -0.013 (0.0080) | 0.00002 (0.00009) | 0.15 | 2.25 | 0.110 | 0.04 (0 – 0.07) |
| **AP gait instability (LDS, λ)** | 100 | 1.14 (0.13) | -0.004 (0.0060) | 0.00006 (0.00007) | 0.12 | 1.06 | 0.350 | 0 (0 – 0.07) |

*Table 2. Effect of age on gait parameters: regression results. The 100 participants walked for 5 min on a treadmill at preferred walking speed. Trunk acceleration in the mediolateral (ML), vertical (V), and anteroposterior (AP) directions were recorded by a 3D accelerometer. Walk ratio was defined as step length divided by step frequency. Gait variability was the RMS of the lateral acceleration, normalized (RMS$_{RATIO}$) to attenuate the influence of speed (see Methods section). Gait instability (local dynamic stability, LDS) was computed using the maximal finite-time Lyapunov exponent method. Regressions (ŷ=Intercept + Slope·age + Quadratic term·age$^2$) were performed after removing outliers (n=sample size). Standard*

*errors are presented parenthetically. RMSE indicates root mean square error. ANOVA results (H0: constant model) are also presented. $\rho^2$ is the unbiased estimate of the coefficient of determination ($R^2$) with the associated 95% confidence interval.*

# Figures

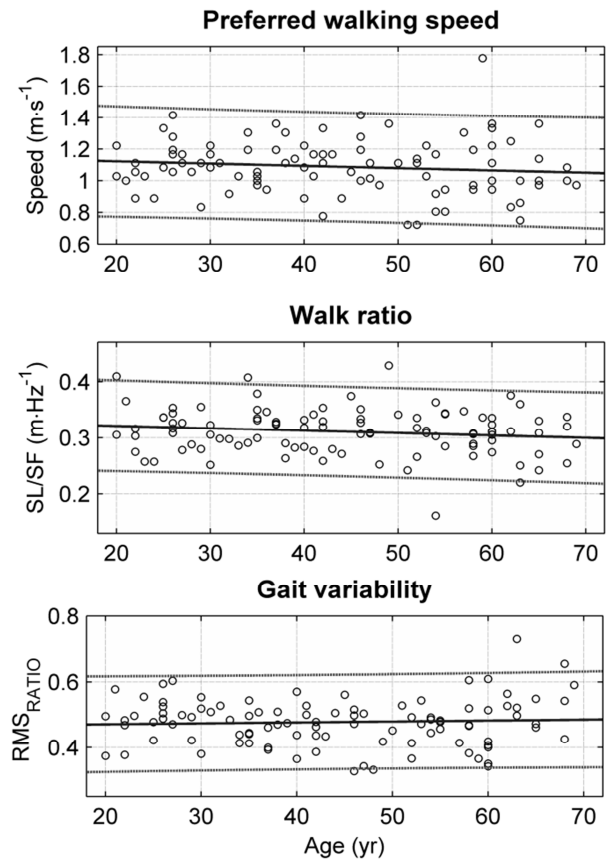

*Figure 1. Scatter plots and regression analyses, spatiotemporal parameters and gait variability.* The 100 participants walked for 5 min on a treadmill at preferred walking speed (top panel). Trunk acceleration was recorded by a 3D accelerometer. Walk ratio was defined as step length divided by step frequency (middle panel). Gait variability was the RMS of the lateral acceleration, normalized ($RMS_{RATIO}$) to attenuate the influence of speed (see Methods section). The value of the dependent variable (y-axis) at the age of each participant (x-axis) is indicated by a small circle (n=100). Best-fit curves resulting from the regression analyses (Table 2) are shown with 95% CIs (±1.96 x RMSE).

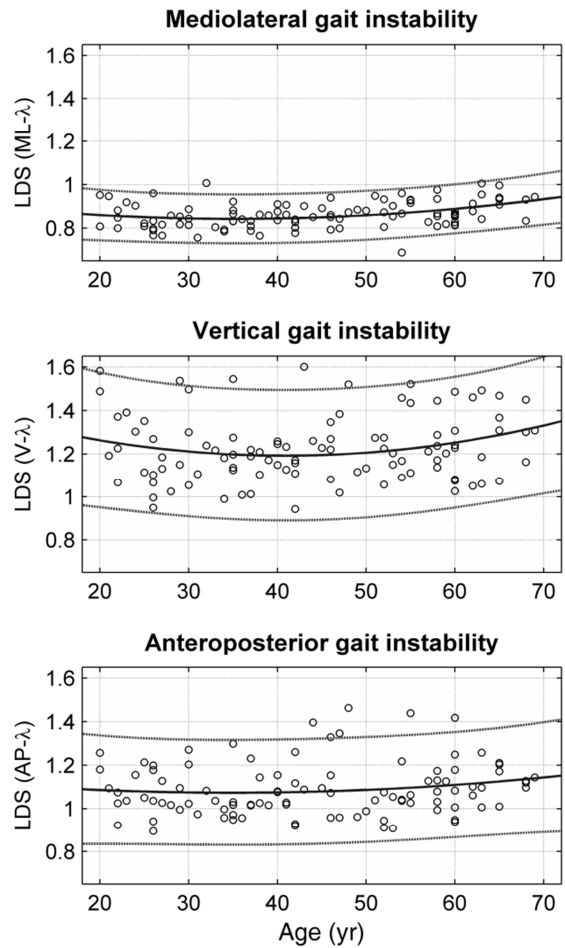

*Figure 2. Scatter plots and regression analyses, gait stability.* *The 100 participants walked for 5 min on a treadmill at PWS. Trunk acceleration was recorded by a 3D accelerometer. Gait instability (local dynamic stability, LDS, λ) was computed using the maximal finite-time Lyapunov exponent method in the mediolateral (top), vertical (middle), and anteroposterior (bottom) directions. The value of the dependent variable (y-axis) at age of each participant (x-axis) is indicated by a small circle (n=100). Best-fit curves resulting from the regression analyses (Table 2) are shown with 95% CIs (±1.96 x RMSE). The 3 panels are presented with identical scales.*